# Network analysis and disease subnets for the SARS-CoV-2/Human interactome

Beatriz Luna-Olivera[1], Marcelino Ramírez-Ibáñez[2,*] and Edgardo Galán-Vásquez[3,*]

[1]UPN unidad 201-Centro de Altos Estudios de la Mixteca, CALMIX, Oaxaca, México. [2]CONACyT-UPN unidad 201, camino a la Zanjita, Nochebuena, 71230, Oaxaca de Juárez, Oaxaca. [3]Departamento de Ingeniería de Sistemas Computacionales y Automatización, Instituto de Investigación en Matemáticas Aplicadas y en Sistemas. Universidad Nacional Autónoma de México. Ciudad Universitaria, México

**Abstract**
**Motivation:** With the aim to amplify and make sense of interactions of virus-human proteins in the case of SARS-CoV-2, we performed a structural analysis of the network of protein interactions obtained from the integration of three sources: 1) proteins of virus SARS-CoV-2, 2)physical interactions between SARS-CoV-2 and human proteins, 3) known interactions of these human proteins between them and the dossier of affections in which these proteins are implicated.
**Results:** As a product of this research, we present two networks, one from the interactions virus-host, and the other restricted to host-host, the last one is not usually considered for network analysis. We identified the most important proteins in both networks, those that have the maximal value of calculated invariants, these proteins are considered as the most: affected, connected or those that best monitor the flow of information in the network, among them we find UBC, a human protein related with ubiquination, linked with different stages of coronavirus disease, and ORF7A a virus protein that induces apoptosis in infected cells, associated with virion tethering. Using the constructed networks, we establish the more significant diseases corresponding with human proteins and their connections with other proteins. It is relevant that the identified diseases coincide with comorbidities, particularly the subnetwork of diabetes involves a great quantity of virus and human proteins (56%) and interactions (60%), this could explain the effect of this condition as an important cause of disease complications.
**Availability:** Network data and programs in Python are included as additional material.
**Contact:** edgardo.galan@iimas.unam.mx or mramirezi@conacyt.mx

## 1 Introduction

Coronaviridae (CoVs) are single-stranded RNA viruses that cause respiratory, enteric, hepatic and neurological diseases of varying severity in a broad range of hosts (Woo et al. 2009). They have the ability to transmit from animals-to-human and human-to-human, as in the case of the several acute respiratory syndrome (SARS) in 2002/2003 and the Middle East respiratory syndrome (MERS) in 2012 (Cauchemez et al. 2013, Cui et al. 2019).

CoVs are subdivided in four genera, where, a) Alphacoronavirus are responsible for gastrointestinal disorders in human, dogs, pigs, and cats; b) Betacoronavirus including the Bat coronavirus (BCoV), the human severe acute respiratory syndrome (SARS) virus and the Middle Eastern respiratory syndrome (MERS) virus; (c) Gammacoronavirus, which infect avian species; and d) Deltacoronavirus, which infect avian and pig species (Woo et al. 2012).

In December 2019, a novel Betacoronavirus (2019-nCoV) was identified from the province of Wuhan, China (Lu et al 2020). It has been causally linked to severe respiratory infections in humans. At time of writing, 26.7 million cases of 2019-nCov have been reported in the World and there are currently 876'616 deaths linked to this pathogen (source: World Health Organization report, 06 September 2020). Phylogenetic relationships suggest that the origin of this new coronavirus is the pangolin (Xiaolu et al. 2020). Although there are advances in the development of a vaccine for 2019-nCoV, there are some challenges to face, including massive testing and large-scale production. Thus, in order to continue to understand how the virus works, we exhaustively evaluated its interactome.

Last studies that are being addressed include topics like: the HCoV–host interactome and drug targets in the human protein–protein interaction network (Zhou et al. 2020), comprehensive structural genomics and interactomics roadmaps of SARS-CoV-2 to infer the possible functional differences and similarities with the associated SARS coronavirus (Srinivasan et al. 2020), the search of master regulators of the SARS-CoV-2/Human interactome (Guzzi et al. 2020). These approaches have used networks as an adequate mathematical approximation to represent and to study several aspects of the same problem. Network analysis has its basis in graph theory (Pavlopoulos 2011), graphs are mathematical objects composed of vertices and edges, or arrows when network is directed. This approach has been associated with several study areas, between them systems biology and cell signaling research. To study the topological aspects of networks, incorporating information about the general and specific properties of vertices, edges, modules, cycles, paths, relevant elements and other components within the network, gives insights into the governing fundamentals of the biological systems and captures their complexity (Ma'ayan 2011).

In this study, we compile the network of SARS-CoV-2/human interactome reported in several papers and databases, and interlace them with the information of the human protein interactions and the analysis of the diseases in which these proteins are involved. In our case, the network is considered as undirected, where every vertex corresponds to a protein, either virus or human and each edge means a physical interaction between SARSCoV-2 and human proteins or human-human proteins. By one hand we identified important vertices from the point of view of network structure using different approaches from network theory, also concepts of connectivity and cointeraction can give insights about essential interactions and proteins, and by other hand we constructed and analyzed the subnetwork of human proteins involved in diseases and their relations with other human and virus proteins.

We consider this study will contribute to prospect studies to increase our knowledge on the interactions of SARSCoV-2 - humans. In follow , we describe the methods and computations that were done in order to get the topological information from the network. Section 3 is devoted to results. Finally, we present the conclusions and discussion of this study in.

## 2 Methods

### 2.1 Interactome network for SARS-CoV-2/Human Cell interaction

We use two complementary types of data for this study. A network was constructed for SARS-CoV-2/human interactome and another complementary network was constructed for the human/human interactome corresponding to the interactions between human proteins involved with the virus proteins, next we describe the characteristics and information sources of both.

SARS-CoV-2 proteins were considered as in (Yoshimoto 2020), taking into account the expressed proteins by the genes: ORF1ab (NSP1 to NSP16), ORF2 (S: Spike glycoprotein or Surface glycoprotein), ORF3a, ORF4 (E: Envelope small membrane protein), ORF5 (M: Membrane glycoprotein), ORF6, ORF7a, ORF7b, ORF8, ORF9 (N: Nucleocapsid phosphoprotein or Nucleoprotein), and ORF10. We add ORF9b because this protein blocks interferon, a key molecule in defense against viruses, Type I IFN antagonist (Estrada 2020) and ORF14, since according to the Multiple Sequence Alignment, ORF14 of SARS-CoV-2 have 77.14% identity with Human-SARS and 92.86% identity with Bat CoV (Wu et al. 2020) .

SARS-CoV-2/human interactome was constructed considering 206 interactions virus/virus and virus/host. It contains 33 virus proteins, 6 of which do not interact with human proteins. In brief, interactions were determined from characterizing viral, intra-viral and human–virus protein complexes, extracting the information on their interaction interfaces and ligand-binding, and superposed the evolutionary difference and conservation information with the binding information (Srinivasan et al 2020). This information was complemented with the network described in BioGRID (Oughtred 2019), which contains 362 virus/virus and virus/host interactions. This database reports 31 virus proteins of which 5 do not interact with human proteins (Gordon 2020, Wang K 2020, Bestle D 2020). In addition, 294 interactions of SARS-CoV-2-Human Protein-Protein network, with 26 viral proteins, which was also used to assemble the network were obtained from VirHostNet 2.0 (Guirimand 2015, Gordon et. al 2020). In VirHostNet, interactions were obtained cloning, tagging and expressing 26 of the 29 SARS-CoV-2 proteins in human cells and identifying the human proteins physically associated using affinity-purification mass spectrometry (AP-MS), Finally, we considered the information from ViralZone, that presents 14 virus proteins without interactions (Hulo 2011). We add two important products of genes with unknown interaction: CTSB, CTSL reported in The Human Protein Atlas (Uhlén 2015).

Therefore, considering the information retrieved from diverse databases, we proceed to construct the interactome network between 404 human proteins involved in interactions with SARS Cov-2 proteins.

Human/human interactome was constructed from diverse sources, String-db provided an interactome human/human network with 1088 protein-protein interactions, in a search of multiple Proteins by Names / Identifiers (Szklarczyk 2019). PICKLE (Gioutlakis 2017) provides a network with 198 interactions in a search of interactions between the selected interactors, their information is obtained from other databases (BioGRID, MINT, HPRD, IntAct and DIP), information about interaction type and the detection method is also reported, for instance: pull down, enzymatic study, AP-MS, biochemical, cross linking study, protein kinase assay, and x-ray crystallography. We performed a random search in HuRI(Luck 2020), which returns 238 interactions between the human proteins affected by the virus. Finally, the resulting interactome includes 432 proteins and 1792 interactions, 404 are human proteins and 28 are virus proteins.

### 2.2 Structural analysis

We consider that every vertex $v$ has a degree $k(v)$, calculated as the number of incident edges on the vertex. The degree distribution was calculated taking into account the relative frequency of appearance for every degree. To deal computationally with a network some programs like Octave (Eaton 2015) use the adjacency matrix $B$, which is constructed with inputs $b_{uv}$ taking $b_{uv}=1$ and $b_{vu}=1$ if there is an edge between $v$ and $u$ and 0 in other cases.

Connectivity in graph theory includes paths, cycles and connected components. A path is a sequence of vertices and edges on the network, such that the initial vertex of the sequence is not the same as the end vertex, and no edges neither vertices are allowed to be duplicated. A cycle is a closed path where the initial and final vertices are the same, the length of the cycle is given by the number of edges in it. In this regard, a graph is connected if there is a path between any two vertices. A network can be composed of several connected components bearing there are groups of proteins that interact separately in the network, a measure which could be suitable in the search of biological groups. In the section of results, we describe the connected components for SARS-CoV-2/Human interactome network.

Clustering coefficient was also calculated for the SARS-CoV-2/human interactome network, this invariant helps us to measure for each vertex if their neighboring vertices are also neighbors between them, that is, it measures the transitivity. For every vertex $v$ the clustering coefficient of the vertex is given by the formula

$$C_v = \frac{2E_v}{(K(v))(K(v)-1)} \quad (4)$$

where $E_v$ is the number of edges between the neighbors of $v$ (Junker 2008).

The simplest centrality is the degree centrality (DC), which gives for every vertex $v$ a measure of the relative connectivity of a vertex in the network (Junker and Schreiber), it is calculated as the degree of the vertex over $n$-$1$, this is the maximum possible degree in a network with $n$ vertices. Other centralities considered in this study are divided in two groups: eigenvector centrality and pagerank centrality, associated with the eigenvalues and eigenvectors of the adjacency matrix, on the other hand we utilize closeness centrality and betweenness centrality which consider the shortest path distance from a vertex to other vertices.

Eigenvector centrality designates the significance of a vertex proportionally to the importance of their neighbors, then a vertex is significant because it is connected to many vertices or because it is connected to vertices with large eigenvector centralities. The value of this centrality in each vertex $v$ is obtained by using the adjacency matrix $B$, their largest eigenvalue $\lambda_n$ and an initial vector $x_v(0)$, we get the value in $v$ by the iteration of function

$$x_v(t+1) = \frac{1}{\lambda_n} \sum_u B x_v(t) \quad (1)$$

The sum is over all the vertices $u$ in the network. A connected network ensures that we can obtain a fixed value $x_v$ after a finite number of iterations. A usual $x(0)$ in computational algorithms is the eigenvector associated to $\lambda_n$.

Pagerank centrality is calculated similarly, with the difference that in the previous sum $\frac{1}{\lambda_n}$ and $B$ are substituted by a weighted matrix $C$, where the value of the input $c_{uv}$ is given by $\frac{B_{uv}}{K_{out}(u)}$ (Langville 2005).

Closeness centrality of a vertex $v$ is defined as the reciprocal of the sum of the length of the shortest paths between the vertex $v$ and all other vertices $u$ in the graph, it is calculated as

$$C_{clo}(v) = \frac{n-1}{\sum_{v=1}^{n-1} d(u,v)} \quad (2)$$

where $d(u,v)$ is the shortest-path distance between $v$ and $u$, and $n$ is the number of vertices in the network.

The betweenness centrality of a vertex $v$ is the sum of the fraction of all-pairs shortest paths that pass through $v$, it is calculated as

$$C_{Bet}(v) = \sum_{s,t \in V} \frac{\sigma(s,t|v)}{\sigma(s,t)} \quad (3)$$

where $\sigma(s,t|v)$ denote the number of shortest paths between $s$ and $t$ that use $v$ as an interior vertex, and $\sigma(s,t)$ is the total number of shortest paths between $s$ and $t$.

Finally, in this study we perform the calculus of vertex energy as a measure of importance. For any matrix $C$, its trace is denoted by $Tr(C)$ and its absolute value $(CC^*)$ by $|M|$. The energy of the graph $G$ is given by

$$\xi(G) = Tr(|B|) = \sum_{i=1}^{n} |B| \quad (5)$$

for a good reference of the energy of a graph see (Li 2012), and for recent applications see (Gutman 2020). For a vertex $v_i$, its energy is defined (Arizmendi 2018) by

$$\xi_G(v_i) = |B|_{ii,} \quad (6)$$

and there are not known applications.

### 2.3 Enrichment analysis

To identify the enriched disease for human proteins, we used the Database for Annotation, Visualization and Integrated Discovery (DAVID; http://david.abcc.ncifcrf.gov/), which is a gene functional classification system that integrates a set of functional annotation tools (Huang et al. 2008). From this, we found 11 groups of diseases, and 105 affected proteins (See additional material), we identified cliques of proteins involved in groups of diseases and for every clique of proteins we prospect the first neighbors proteins, that are divided in human or virus proteins, this first neighbors proteins which could also be related to these diseases. For every group of diseases we constructed a subnetwork which contains: the clique of proteins identified by enrichment, first human and virus neighbors and their interactions observed in VHN, this permits us to perceive the effect of proteins involved in diseases.

### 2.4 Algorithms and implementation

Algorithms were implemented in Networkx from Python (Rossum 2020). The following routines are already included in Networkx: Degree, Eigenvector centrality, connected components, Katz centrality, Closeness centrality and Betweenness centrality. In additional material we provide the algorithms to find: degree distribution, and vertex energy. As a negative control we construct random Erdos-Renyi networks with the same number of vertices and interactions to contrast the results of invariants.

## 3 Results

### 3.1 Degree, degree distribution and connectivity

The resulting interactome including human/human and virus/human interactions proteins consists of 432 proteins and 1792 interactions, this is the virus/human network (VHN), of these edges 123 are auto interactions, that is to say around 7%, it is organized in only one connected component.

The interactome of the proteins involved with the virus is of 404 proteins and 1247 interactions and is represented in a human/human interactome network (HHN). It contains 107 auto interactions, this is roughly 8.5%, and it is organized in one giant connected component with 299 proteins (on which the following calculations are presented), 2 connected components of size 2, and the rest of vertices are isolated or auto interacting, meaning at this level are the virus proteins that unite the network.

In biological networks, degree and degree distribution are outstanding because they help to elucidate the global structure of the network. In this context, we identified that excluding the auto interactions, for VHN the vertex with maximum degree is ORF7A with degree 60, followed by NSP13 with degree 52 and NSP7 and UBC with degree 51, the minimum degree is 1 and 92 vertices reach this degree, the mean is 7.7, variance 67 and median 5. The adjustment of degree distribution is given by the equation $y = 200x^{-1.35}$ (Figure 1A).

In HHN the vertex with maximum degree is UBC with degree 50, followed by HNRNPA1 with degree 30, and DDX5 with degree 27. The mean is 7.6, variance 39, and median 6. The adjustment of degree distribution is given by the equation $y = 113.78x^{-1.18}$ (Figure 1B).

### 3.2 Clustering coefficient

We found that the highest clustering coefficient in both networks is 1, *i.e.* vertices whose neighbors are all connected between them forming complete graphs denoted by $K_n$ for $n$ vertices.

For VHN clustering coefficient equal to 1 was found for BCL2A1 with degree 4; and for AASS, TPSAB1, TPSB2, KIAA1033, DCAKD and CCDC86 which possess a degree of 2, meaning we found triangles. Otherwise, 92 vertices has a clustering coefficient equal to 0, and degree 1, which corresponds to 21% of vertices in the network; whereas 66 vertices have different degrees, from 2 to 6, and clustering coefficient of 0, these are called stars in graph theory, and are subgraphs composed of one central vertex and its neighbors, being the most notorious the proteins NSP5, MKRN3, PPT1 and PKHF2. The mean of clustering coefficient for the network is 0.19, with a variance of 0.05, and median 0.13. This average indicates that neighbors have 1/5 of connections they could have.

For HHN, we found in 9 vertices, a clustering coefficient of 1, for TRM1 with degree 7, meaning $K_8$ is present in this network, NUP58 with degree 5 give place to $K_6$. We obtain $K_5$ with TIM29, PGES2, BCL2A1, NGLY1 of degree 3. We also found vertices with different degrees and clustering coefficient 0, being the most notorious the protein HMOX1 with 7 neighbors, and EMC1, SGTA, DPH5 with 6. The mean of clustering coefficient for the network is 0.24, with a variance of 0.05 and median 0.2, this average indicates that neighbors have 1/4 of connections they could have. Vertices with clustering coefficients equal to 1 are radically different in VHN and HHN, meaning that interactions with virus proteins affect the original complete pieces in HHN introducing new interactions and creating new complete graphs with other proteins. The clustering coefficient by degree is presented in Figure 1C and 1D.

### 3.3 Centralities

Centralities for the VHN network are compared in Table 1, we present the top ten vertices with the highest centrality values in each case. As we can see, some elements are consistently repeated in several centrality measures, for instance human protein UBC, and virus protein NSP8 are identified in all centralities. Followed by NSP7, NSP13, ORF7A and NSP12 all virus proteins. Next in frequency are: DDX5, HNRPA1 from humans and M and ORF8 from viruses.

**Table 1.** Centralities of VHN.

| Level | Degree | Betweenness | Closeness | Eigenvector | Vertex energy |
|---|---|---|---|---|---|
| 1 | ORF7A | NSP7 | NSP7 | UBC | NSP13 |
| 2 | NSP13 | NSP8 | NSP8 | HNRPA1 | NSP7 |
| 3 | UBC | NSP13 | ORF7A | DDX5 | ORF7A |
| 4 | NSP7 | ORF7A | NSP12 | RPS20 | UBC |
| 5 | ORF8 | NSP12 | ORF3A | EEF1A1 | NSP8 |
| 6 | NSP8 | M | ORF9B | PABP1 | M |
| 7 | M | ORF3A | NSP13 | DDX10 | NSP12 |
| 8 | NSP12 | UBC | UBC | U3IP2 | ORF8 |
| 9 | HNRPA1 | NSP9 | E | NSP8 | NSP9 |
| 10 | DDX5 | ORF8 | DDX5 | POLR2B | HNRPA1 |

The top ten proteins identified by every measure of importance in VHN.

Centralities for the HHN network are compared in Table 2, we present the top ten vertices with the highest centrality values in each case. As we can see, some elements are consistently repeated in several centrality measures, for instance: UBC, HNRPA1, DDX5 and POLR2B. Followed by UBE2I, and DDX10.

**Table 2.** Centralities of HHN

| Level | Degree | Betweenness | Closeness | Eigenvector | Vertex energy |
|---|---|---|---|---|---|
| 1 | UBC | UBC | UBC | UBC | UBC |
| 2 | HNRNPA1 | RAB1A | HNRNPA1 | HNRNPA1 | HNRNPA1 |
| 3 | DDX5 | HNRNPA1 | EEF1A1 | DDX5 | RAB7A |
| 4 | RAB7A | RAB7A | UBE2I | RPS20 | DDX5 |
| 5 | DDX10 | UBE2I | DDX5 | PAPB1 | POLR2B |
| 6 | PABP1 | DDX5 | RAB1A | EEF1A1 | UBE2I |
| 7 | EEF1A1 | GOLGA2 | RPS20 | DDX10 | GOLGA2 |
| 8 | POLR2B | POLR2B | POLR2B | U3IP2 | PABP1 |
| 9 | UBE2I | OS9 | PSMA2 | POLR2B | RAB1A |
| 10 | RPS20 | ARF6 | GOLGA2 | MPP10 | DDX10 |

The top ten proteins identified by every measure of importance in VHN.

To evaluate the structure of the networks with respect to chance, we reconstruct random networks with the same number of nodes and interactions. We found that random networks are significantly different from real. Degree and eigenvector centralities present a normal distribution and lower values, as well as vertex energy. Contrary to what happened in VHN and HHN, in random networks there are no vertices with clustering coefficients equal to one and there is a big difference between the values of betweenness centrality. Notoriously closeness centrality does not have variation in its distribution, meaning is not an invariant to allow distinguish between VHN, HHN and random networks of type Erdos-Renyi.

### 3.4 Diseases proteins

From the enrichment analysis we found eleven cliques of proteins related with diseases, each group of diseases is shown in the first column of Table 3 and the number of proteins taking part in it is mentioned in the second column. The number of first neighbors of every clique that are human proteins is mentioned as NHP and the number of first neighbors of every clique that are virus proteins is mentioned as NVP, the number of interactions in the subnet is reported as NI. The same information can be visualized in Fig. 2.

From the enrichment analysis we found that the human proteins BCL2, GPX1, RIPK1, VKGC and VKORC1 are involved in diverse diseases (4).

From the neighbor search we detect 23 virus proteins interacting with proteins involved in diseases, the most common is ORF7A present in 9 groups of diseases, excluding only: Severe Acute Respiratory Syndrome and HIV Infections|Sexually Transmitted Diseases. There are 261 human proteins related to disease proteins, the most common are: COMT, PPIG and VKORC1, which are involved in 7 diseases at the same time. At the structural level, these proteins play as links between diseases, which allows a completely connected structure of the network.

Examining every subnetwork by disease, we found that the most extended contained in VHN is the subnetwork corresponding to: Type 2 Diabetes, edema and rosiglitazone. There are 100 human proteins that appear in this subnetwork but are not related with any other disease. NSP1, NSP2 and NSP3 are viral proteins exclusively involved in Diabetes.

ORF6 is exclusive for Acquired Immunodeficiency Syndrome, Disease Progression. ACE2 is identified as an important protein involved in Acquired Immunodeficiency, Type 2 Diabetes| edema | rosiglitazone, Colorectal Cancer, Severe Acute Respiratory Syndrome, HIV Infections|Sexually Transmitted Diseases. The three last diseases in Table 3 share the same vertices and thus the same virus and human neighbors.

**Table 3.** We present the groups of diseases found in enrichment analysis, the number of proteins involved in diseases (NP), the number of first neighbors of this proteins clique, which could be human (NHP) or virus proteins (NVP) and the final number of interactions (NI) in the sunetwork corresponding to the disease group.

| Disease | NP | NVP | NHP | NI |
|---|---|---|---|---|
| Acquired Immunodeficiency Syndrome|Disease Progression | 40 | 14 | 67 | 298 |
| Type 2 Diabetes| edema | rosiglitazone | 60 | 21 | 161 | 1074 |
| Colorectal Cancer | 18 | 12 | 78 | 391 |
| Lymphoma, B-Cell|Lymphoma, Follicular|Lymphoma, Large B-Cell, Diffuse | 5 | 3 | 20 | 65 |
| Hodgkin Disease|Leukemia, Lymphocytic, Chronic, B-Cell|Lymphoproliferative Disorders|Waldenstrom Macroglobulinemia | 7 | 5 | 25 | 83 |
| Severe Acute Respiratory Syndrome | 4 | 2 | 5 | 14 |
| HIV Infections|Sexually Transmitted Diseases | 2 | 1 | 4 | 11 |
| Venous Thromboembolism | 3 | 3 | 10 | 20 |
| Apoplexy|Atherosclerosis|Stroke | 2 | 2 | 4 | 9 |
| Warfarin therapy, response to | 2 | 2 | 4 | 9 |
| Protein C Protein S | 2 | 2 | 4 | 9 |

The eleven diseases enriched in the VHH network found using DAVID (Huang et al. 2008).

## 4 Discussion

In this work we integrate the SARS-CoV-2/human interactome and interlace them with the information of the human protein interactions and the analysis of the diseases in which these proteins are involved.

The first topological invariant we observe are degrees, related in some works with level of expression and indispensability (Estrada 2006, Giménez 2016), in this case the most connected protein in the VHN is ORF7A, it interacts with 60 other proteins that include 5 viral proteins and 55 human proteins. ORF7A directly binds to BST-2 (also known as CD317 or tetherin) and inhibits its activity by blocking the glycosylation of BST-2 (Taylor et al 2015), in addition ORF7a can induces apoptosis in infected cells, contributing to the damage in the lugs (Yeung 2016).

Another highly connected node is NSP13 that interacts with other 52 proteins. NSP13 It is a helicase that unpacks viral genome material to make it more accessible, it adopts a triangular pyramid shape comprising five domains (Mirza and Froeyen 2020).

The most important protein in the HHN network is UBC, which is at the same time the most connected protein that interacts with other 50 human proteins. It is a polyubiquitin-C and it plays a key role in maintaining cellular ubiquitin levels under stress conditions.

Another important node in the HHN network due to its connectivity and centrality is HNRNPA1, it is a heterogeneous ribonucleoprotein A1, it is targeted by viruses to control various stages in their life cycle: replication, transcription, post transcriptional modification like nuclear export/import, and translation (Kaur and Lal 2020) and DDX5 is a RNA binding protein, it plays multifunctional roles and is involved in all aspects of RNA metabolism (Jalal et al 2007).

Difference of degree between VHN and HHN provides information about the human proteins more affected by virus proteins, which in this case are the proteins: SERPING1 with 6 new interactions, DCTN2 with 4 new interactions and PPIA with 3 new interactions. SERPING1 is a C1 inhibitor, a type of serine protease inhibitor (serpin). Serpins help control several types of chemical reactions by blocking the activity of certain proteins. C1 inhibitor is important for controlling a range of processes involved in maintaining blood vessels, including inflammation. Inflammation is a normal body response to infection, irritation, or other injury (Yurchenko et al. 2010). PPIA encoded a Cyclophilins A can bind and activate the transmembrane receptor CD147. Cyclophilins play a critical role in the replication process of HIV-1, HCV and many other viruses (Pushkarsky et al. 2001). DCTN2 encodes a Dynactin, Dynactin binds to both microtubules and cytoplasmic dynein. It is involved in a diverse array of cellular functions, including ER-to-Golgi transport, the centripetal movement of lysosomes and endosomes, spindle formation, chromosome movement, nuclear positioning, and axonogenesis.

We calculated centrality measures for vertices into the network in order to find essential proteins, in (Estrada 2006) author indicates that essential proteins generally have more interactions than the nonessential ones and associates lethality with the remotion of proteins with high centralities in the yeast proteome, also indicates that choose essential proteins considering their high centralities is significantly better than random selections.

In the VHN the most central node according to degree centrality is ORF7A. NSP7 and NSP8 are identified as the most central nodes according to Closeness and Betweenness centralities, which implies these proteins are the nodes that are able to spread the information very efficiently through the network and they are the nodes that best monitor the flow of information in the network. NSP7 and NSP8 are cofactors that interact with NSP12 to form the SARS-CoV-2 polymerase complex (Peng et al 2020).

NSP13 has the highest vertex energy, though there are not known applications of this invariant it seems to compile the information of other three centralities: degree, betweenness and closeness centrality in a unique measure. while UBC is the most central according to eigenvector centrality, which indicates that this node is connected to other highly connected nodes.

On the other hand, in the VHH network the most central node in all the centralities is UBC, it is admitted that ubiquitination is related with several important cellular functions like: DNA repair, cell cycle regulation, kinase modification, endocytosis, and regulation of other cell signaling pathways, moreover in (Raaben et al 2010) identified that the ubiquitin proteasome system was involved in several steps of infection of coronavirus.

An important aspect of SARS-CoV-2, is the susceptibility of infection of patients with certain comorbidities that can trigger higher mortality. The most recurrent comorbidities are Cardiovascular diseases, Hypertension, Diabetes, Chronic obstructive pulmonary syndrome, Chronic kidney disease and Cancer (Gold et al 2020, Bajgain et al 2020). In this context, we identified 11 enrichment groups of proteins related with diseases (Fig. 2), we found these groups are related to previously described comorbidities, which indicate that the interactions between the SARS-CoV-2 viral proteins and human proteins are intensified with these comorbidities.

The most abundant group is the related to Type 2 Diabetes| edema | rosiglitazone, this consists of 60 human proteins, as well as, 21 viral proteins. We identified that NSP1, NSP2, and NSP3 are proteins that are exclusively found in diabetes subnetwork. NSP1can inhibit IFN signaling and block the host innate immune response by promotion of cellular degradation and blocks translation of host's RNA (Kamitani et al 2006). NSP2 interacts with a host protein complex of PHB1 and PHB2 involved in mitochondrial biogenesis (Kumar et al 2020), it is the major virulence factor which suppresses host gene expression by binds to 40S and 80S ribosomes (Thoms et al. 2020). NSP3 Promoting cytokine expression and cleavage of viral polyprotein (Astuti 2020).

Another important aspect in the SARS-CoV-2 infection is the formation of Blood clots in the patients (Lemke and Silverman 2020). In this context, four groups of diseases identified are related to effects in the blood: 1) Apoplexy | Atherosclerosis | Stroke that are involved in internal bleeding, blood flow deficiency and arterial blockage; 2) Venous thromboembolism which is a disorder that includes deep vein thrombosis and pulmonary embolism; 3) Warfarin which is a vitamin K antagonist and inhibits synthesis of vitamin K-dependent clotting factors (II, VII, IX, X) and 4) proteins C and S, two proteins in the blood that help regulate blood clot formation.

Particularly in Protein C Protein S disorder, and this affectation may be due to the interactions between ORF7A and VKORC1 which is Vitamin K epoxide reductase complex subunit 1, which is involved in vitamin K metabolism. Vitamin K is required for the gamma-carboxylation of various proteins, including clotting factors, and is required for normal blood coagulation. The interaction of these viral proteins with human proteins may explain the failure at the global level present in Sar-CoV-2 patients.

In conclusion, this study not only offers an integrative network-based system that allows finding important nodes in the interaction of SARS-CoV-2 and the human proteome, which empower the identification of possible markers for response to drugs, but also provides a feasible explanation of comorbidities.

## Acknowledgements
We acknowledge Ernesto Perez-Rueda for their critical comments to the final version of this document.

*Conflict of Interest:* none declared.

*Data availability:* The data underlying this article are available in the article and in its online supplementary material.

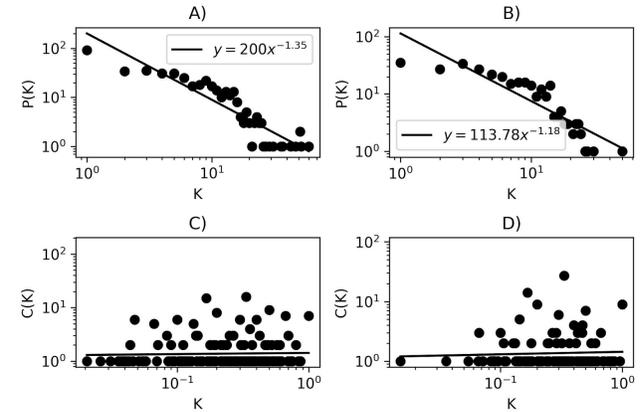

**Fig. 1. Topology of networks.** A) Degree distribution of VHN and B) Degree distribution of HHN, C) Clustering coefficient of VHN, and D) Clustering coefficient of HHN.

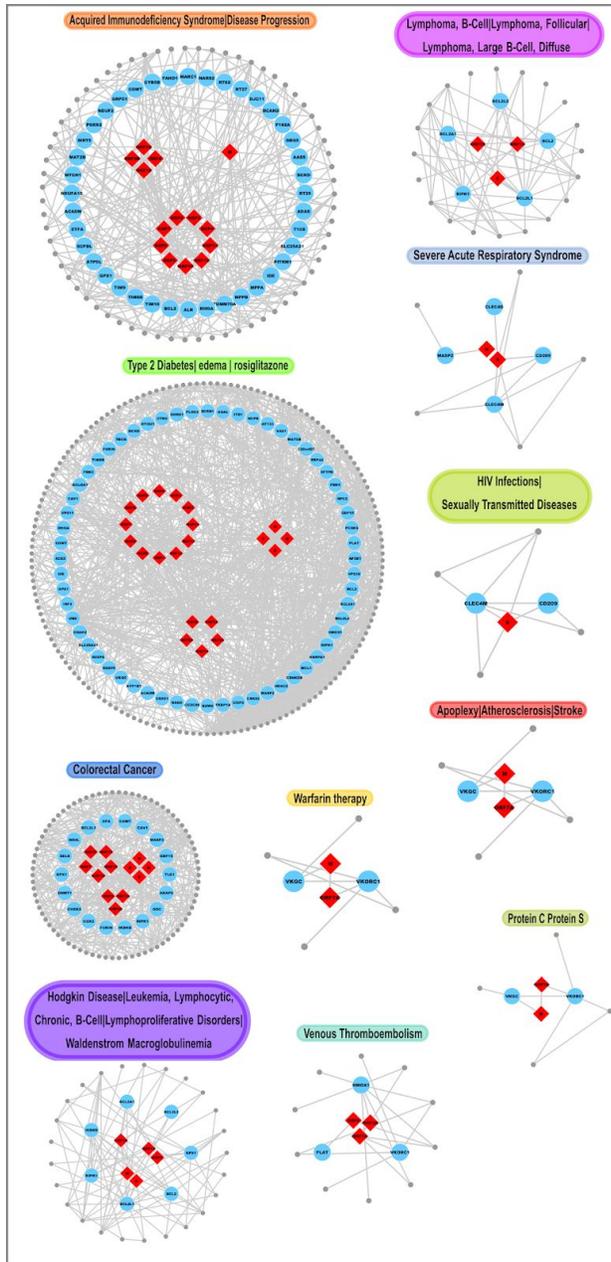

**Fig. 2. Disease subnets enriched in VHN.** Each of the subnets corresponds to a group of diseases and it is represented by three sets of nodes: blue nodes make up the clique of human proteins identified as disease elements in enrichment analysis, red and gray nodes are first neighbors of this clique of proteins, virus and host proteins respectively. Interactions were taken as in VHN.